# Quorum Sensing as a long-range interaction for bacteria growth and bioluminescence

Eleonora Alfinito [1,*], Matteo Beccaria [1,2,3], Maura Cesaria [1]

[1]Dipartimento di matematica e fisica 'Ennio De Giorgi', Università del Salento, Via Arnesano, I-73100 Lecce, Italy

[2]Istituto Nazionale di Fisica Nucleare - sezione di Lecce, Via Arnesano, I-73100 Lecce, Italy

[3]National Biodiversity Future Center, Palermo 90133, Italy.

*eleonora.alfinito@unisalento.it

## Abstract

We study the role of Quorum Sensing (QS) in the growth of bacterial colonies and in the bioluminescence produced. These two phenomena are both regulated by QS and the experimental data show a non-trivial correlation between them. It is also known that the specific bacterial substrate potentially modifies the behavior of the colony. In the specific case of the bioluminescent gram-negative bacterium, *Vibrio harveyi*, we propose a three-autoinducer model in which QS is described in terms of long-range interaction between charged objects placed on a regular network and playing the role of bacteria. The charges spread in the network through fictitious non-linear electrostatic interactions. QS is monitored by analyzing the current flowing within the network. The model parameters are determined by comparing the simulations with the data present in the literature and related to liquid cultures. New colony growth measurements are then performed on *Vibrio campbelli* (a member of the clade *Harveyi*) grown on hydroxyapatite (HA) substrates relevant for biomedical applications. Growth on the substrate differs from growth in liquid culture, although the observed bioluminescence is similar. We exploit our model to understand the differences between colonies in liquid culture and on substrate in terms of the relative role and cooperation of the three autoinducers.

## I. INTRODUCTION

Quorum sensing (QS) describes the cooperative behavior of bacteria in up/down regulation of gene expression. It governs several activities related to their life, like reproduction, biofilm production and development, bioluminescence [1], toxin production and so on [2,1]. QS is a long range phenomenon mediated by some signaling molecules, called autoinducers (AIs), that are secreted by each bacterial cell and become effective in collective communication only when a critical concentration of bacteria is reached. It is very common that bacteria strains, communicate through multiple different types of AIs [3,4,1]. In this case, the AIs act synergically to implement QS [1]. The effects of QS on the regulation of different biological features have not always been clarified nor it is clear whether it always manifests itself in the same way. Measurements of QS are always indirect, *via* its effects, thus making difficult to understand the basic mechanisms of the phenomenon. Experimental and theoretical investigations focused on the analysis of specific QS-induced features, and so far a more complete view of the phenomenon is still lacking. Here, we propose a model of QS which is able to capture more than one feature associated with a well-known marine bacterium, *Vibrio harveyi*. This bacterium is widely investigated because it is pathogenic for different marine species and, perhaps, for humans [5], too. *V. harveyi* implements QS by means of three different autoinducers which are produced and detected in parallel: homoserine lactone (denoted HAI-1), the so-called autoinducer-2 (AI-2), the *V. cholerae* autoinducer-1, termed CAI-1 [1]. The first one is specific of this *specie*, the second is an intraspecies signaler, the last one is specific of the pathogenic bacterium *V. cholerae*, although it plays a main role in the regulation of several activities of *V. harveyi* [1]. *V. harveyi* is a bioluminescent organism with luminescence yield increasing with the density of the colony, and is therefore a measure of QS [1]. The role of the three AIs in the regulation of bioluminescence was studied in [1] in a liquid culture: the response given by a set of seven mutants of *V. harveyi* strains, different for the kind of AIs systems that they express, was analyzed. Bioluminescence was collected for the wild type (which contains the complete set of AIs), three mutants expressing only one of the AIs, (single-component-mutants), and the three different combinations of mutants expressing two AIs (two-component-mutants). Remarkably, it has been observed that the deletion of one or two of the AI systems dramatically reduces the light production [1]. On the other hand, the deletion of one AI can speed up the colony formation [6] for bacteria grown in liquid culture.

In our approach, we explore the effects of the synergic action of the three AIs on colony development and bioluminescence, by means of a model which takes into account the role of AIs as mediators of QS using a long-range potential on a regular network. In this non-linear model the potential grows with the amount of bacteria on the network in a positive feedback loop.

We validate the model on the available data for *V. harveyi* in liquid culture in order to assess the precise role and correlation of the three auto-inducers. A further complication that reveals the complex nature of bacteria colony dynamics is the environment where the colony lives, which may play a role. Because of this reason, as a further test of our model, we performed new experimental measurements on colony growth on a solid substrate, i.e. hydroxyapatite (HA), which is a common biomaterial with important application in biomedicine [7]. Comparing simulations with the novel data in this different setup supports the idea of a specific role of each autoinducer in the regulation of different bacteria features, although in the framework of a cooperative behavior [1].

## II. MODEL

Many different deterministic [8-10], and stochastic [11-13],models have been developed to describe some specific processes regulated by QS, such as biofilm formation [14,15], bioluminescence [9], pattern formation [16] ,motility [17] and so on, often related to specific bacteria families [10] and using different approaches, analytical [8], stochastic [11],computational [15] , finally analyzing also social behaviours related to QS [18].

This paper proposes a stochastic discrete model, previously introduced in [19] and inspired on random resistor network models [20]. The bacteria colony is left to grow on a regular network where each site represents a bacterium/bacteria aggregate. Sites with positive charge are called active. Each site has a strength measured in terms of its charge Q and representing the effectiveness in terms of genic regulation. QS-like behaviors emerge due to the long range interaction potential produced by the charges, which induces a single charge dynamics as a function of the total amount of charges in the network. Furthermore, QS may be quantitatively monitored by computing the impedance of the network [19]. As in real systems in which QS regulates different activities, we propose here at least two different ways in which QS can act, i.e. colony formation and bioluminescence production. As it turns out, the charge of each bacterium evolves in such a way to maximize the energy of the colony, increasing its value and also colonizing the nearest neighbors. The higher the total energy, the higher the probability for each charge to improve its value. Furthermore, each active site may activate a close site of lower-potential by transferring half of its charge .

The process is limited because each charge can grow until a maximum value is reached. Then, it dies and may be replaced due to the colonization of nearby alive charges [19].

The model is divided into 3 modules:

(i) INOCULATION: A regular square grid is set up to describe the substrate on which the colony can develop. Each site represents a bacterial entity with strength identified by an integer charge that produces a long-range interaction with all the other charges of the network by means of the potential:

$$V(r') = \int_{net} \frac{\theta(r)}{|r-r'|} dr , \qquad (1)$$

where $\theta(r)$ is the charge density and integration is performed on the whole network . The energy of each charge is: $e(r') = q(r')V(r')$ and $energy(t) = \int e(r) dr$ represents the whole network energy.

The charge is initially randomly distributed (here, to occupy 5% of the sites ). The potential is recomputed at each step .

(ii) COLONY LIFE. Each active node probes other active nodes among the nearest neighbors, having lower potential (receivers) thus opening channels among them. This happens with probability:

$$P(r,r',t) = \min(1, \exp[-\beta(e(r) - e(r'))]) , \qquad (2a)$$

where:

$$\beta^{-1} = \frac{energy(t)}{\sum a_i} \text{ and } \sum a_i \leq 1. \qquad (2b)$$

In this formula the quantities $a_i$ represent the relative AI significance, i.e. the percentage of total AI produced by the strain.

When a channel is open, the receiver improves its charge according to the rule:

$$Q(n) \to Q(n) + floor(\frac{\sigma * links(n)}{N}) \qquad (3)$$

where $1 < \sigma < N$ is a real number whose value determines the efficiency of activation (here 22), and $links(n)$ is the number of sites connected to the n-th site. The final result consists of the single charge growth . In this step there is no way to reduce the charge value [19].

Each charge explores the nearest neighbor nodes and may colonize one of them, transferring half of its value. In the case of a too small charge (Q=1) migration is allowed towards one of the nearest neighbor sites [19]. In both cases, whether it is colonization or migration, the sites with the lowest potential are preferentially selected, in such a way to easily improve the charge value in the next steps. When the charge reaches the maximum value, $Q_{max}$ (here 80), it cancels ('dead') and is replaced by new charge coming from the neighboring sites. In conclusion, the charge value may change according to the possible transitions that we have described:

$$0 \to mQ, \qquad m \geq 1$$

$$nQ \to \frac{n}{2}Q, \qquad n \geq 2$$

$$Q_{max} \to 0 \qquad . \qquad [4]$$

These dynamics may be regarded as a compartmental model [21] which describes birth, reproduction, and death of the charge.

The model reaches a stationary state characterized by oscillations of frequency dependent on the maximal value (age) of Q: the larger Q the slower the oscillations. These oscillations can be suppressed by choosing to no longer replace a dead bacterium [19].

(iii) QS IMPLEMENTATION. In the model, QS is visualized by means of the flow of electrical current inside the network. The current, injected by means of ideal contacts put on the left and side of the network, goes through the active nodes by means of the channels opened in module (ii). The resistance of these channels varies according to the law:

$$R(n,m) = D(n,m)\left[\begin{array}{c}\rho_{min}h(n,m)\\ +\rho_{Max}(1-h(n,m))\end{array}\right] \quad (5)$$

where $\rho_{min}, \rho_{Max}$ are the minimal/maximal resistivity of the channel, $D(n,m)$ is the channel length, and $h$ is the Hill-like function [22,23]:

$$h(n,m) = \frac{Q(n)^\gamma Q(m)^\gamma}{g^\gamma + Q(n)^\gamma Q(m)^\gamma}, \quad (6)$$

where Q(n) is the charge of site n, and $g, \gamma$ are real numbers. In more details, γ, the Hill number, determines how fast the function tends to 1. We remind that the Hill function is a standard tool for estimation of the affinity of multiple ligands of the same receptor. In that context, the parameter γ is a measure of the number of ligand while g is related to the dissociation constant [23], here g=8 [19].

Eqs. (4,5) describe the sigmoidal trend of the resistance which develops between its maximal ($D\rho_{max}$) and minimal values, $D\rho_{max}$ and $D\rho_{min}$. The total current collected in the out- electrical contact is therefore a measure of the amount of charge (Q) in the network and also of its distribution.

### III. MEASUREMENTS ON HYDROXYAPATITE

Four strains of *V. harveyi* strain ATCC BAA-1116 ( recently reclassified as V. *campbellii* [24]) were investigated experimentally, that is, the wild type strain BB120 and the associated mutants JAF633 (*ΔluxM*—i.e. deprived of HAI-1 ), KM387 (*ΔluxS,* i.e. deprived of AI-2) and JMH603 (*CqsA::Cmr,* i.e. deprived of CAI-1),grown onto hydroxyapatite (HA, $Ca_{10}(PO_4)_8(OH)_2$) disc-like substrates. Inertness makes this material widely used in the biomedical field. In this context, it is very suitable to compare the predictions of the model with experiments due to its negligible impact on the development of the mutant-dependent growth of the biofilm.

The samples were prepared and grown according to the protocol detailed elsewhere [25]. Hydroxyapatite substrates were placed in sterile multi-wells and exposed to 1 mL of planktonic solution for 24 and 48 h at 37 °C. Following the removal of the agar medium broth, the samples were washed for three times by immersion in a sterile saline solution (0.9 % NaCl) and bacteria were fixed by means of immersion in methanol (100 %) for 15 minutes. The staining of the biomass developed onto HA discs was obtained by treating the samples for 5 minutes with a solution of CV (crystal violet) diluted with 8 volumes of water [26]. Subsequently, the CV-stained samples were analyzed by a stereomicroscope (Nikon, SMZ 1270, zoom ratio 12.7:1, zooming range 0.63-8x) to image the characteristics (morphology and coverage) of the biofilm distribution onto HA. Quantitative estimation of the biomass was performed by optical density measurements at 595 nm (in brief, OD (595 nm)) of the ethanol-based whitening solution following growth for 48 hours and by digital counting of colonies (alive bacteria) yielding the percentage values of Colony Forming Unit (CFU) in terms of CFU/mL [25]. The OD (595 nm) signal is proportional to the whole biomass grown on the supports that is detached by ethanol and dispersed in a solution. Moreover, thresholding and segmentation of the stereomicroscopy images by the license-free software ImageJ [27] allowed to estimate the relative percentage coverage associated with each mutant (KM387, JAF633 and JMH603) calculated according to the following formula :

$$cvg_{rel}(Mutant) = \frac{|\%A(Mutant) - \%A(BB120)|}{\%A(BB120)} \quad (7)$$

where %A(Mutant) refers to the percentage area associated with any mutant (i.e., Mutant= KM387, JAF633, JMH603) and %A(BB120) is the percentage area of the wild type strain BB120. The quantity expresses the change in the coverage of the mutant relative to BB120 reference strain.

### IV. RESULTS

We have compared the prediction from the above model with three specific sets of experimental data, looking for correlations between QS, bioluminescence and colony growth.
1. Bioluminescence in liquid culture. Experiments in [1] report on bioluminescence data obtained at high cell density, measured on the wild type strain and six different mutants.
2. Colony growth in liquid culture. Experiments in [6] report on the growth rates of the wild type strain and several different mutants .
3. Colony growth on the hydroxyapatite substrate. Experiments in this paper report on the growth rates of the wild type strain and three isogenic mutants .

### A. QS and bioluminescence in liquid culture

In [1] bioluminescence measurements were performed on different mutants of *V. harveyi*. The highest luminescence is produced by the wild type, (BB120), containing the whole set of AIs. Mutants deprived of a single AI (AI⁻) produce luminescence in the following order: CAI-1⁻ >AI-2⁻>HAI-1⁻ [1]. Single- component mutants are dim or dark [1]. The two-component mutants show a reduction in light of about 83%, 96% and 99%, respectively. This shows that bioluminescence is effective only in the wild type and, furthermore, assigns to CAI-1 a marginal role in its production. On the other hand, although marginal, CAI-1 is necessary to obtain the maximal response (wild type). Mutants deprived of HAI-1 give a poor response in terms of bioluminescence although better than the one given by mutant deprived of HAI-1 and also of CAI-1. In other words, a specific cooperativity of each AI, different in the presence of none, one or two different AIs is postulated. Light intensity covers about 3 orders of magnitude; therefore it is possible to discriminate among the responses with quite good accuracy.

Bioluminescence provides an effective way to visualize QS and thus module (iii) of our model is used to fit experimental data. Specifically, the asymptotic mean value of the total current flowing in the network is here assumed as a bioluminescence analogue (BA). The final value of BA should accounts for cooperativity by means of the value of γ (see Eq.6). Simulations were performed using grids of different sizes, specifically 40x40, 50x50 and 60x60.

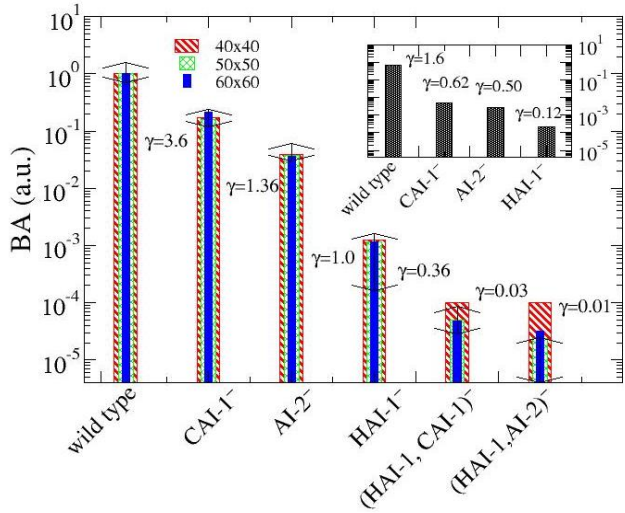

FIG. 1. Bioluminescence analogue (BA) calculated for different grid sizes. The bar errors reproduce the experimental data obtained in [1]. Inset: BA calculated for the samples grown on HA. All data are normalized to the maximal value obtained for the wild type BA is normalized to the same value reported in the main figure.

When the bioluminescence of two-component-mutants is analyzed, the effective γ value is found in the range 0.36-1.36, i.e. from a non-cooperativity to a modest cooperativity value. On the other side, each of the AIs in the presence of the other two AIs further enhances its cooperativity and the maximal bioluminescence is obtained using γ= 3.6, a significantly larger value than 1.36 thus suggesting a net cooperation among the AIs.

In Figure 1 we report the BA normalized to the maximal value obtained for each grid size. Analyses performed using different grid sizes are in good agreement with experiments for the wild type and the three double-component mutants. Concerning the single-component mutants, their BA should be obtained using very small values of γ ($10^{-2}$ - $3\times10^{-2}$) which produce very noisy data. Finally, the experimental response of the single component mutant containing only HAI-1 is below the error and has not been analyzed in this paper.

### B. QS and colony growth in liquid culture

Modules (i) and (ii) are responsible for the colony growth which depends on the amount of AI ($a_i$ coefficients), see Eq.(2a).

Figure 2 reports the running average of the total charge in the network, calculated for different combinations of AIs, i.e. different mutants. Calculations were performed by using the following correspondence $a_1 = 0.1$ for CAI-1, $a_2 = 0.3$ for AI-2 and $a_3 = 0.6$ for HAI-1, which well fits data reported in the literature [6].

Our model describes a behavior in which the colony, after a fast rise, approaches a dynamical equilibrium ruled by equations (3), i.e. characterized by global charge oscillations due to the death-birth processes.

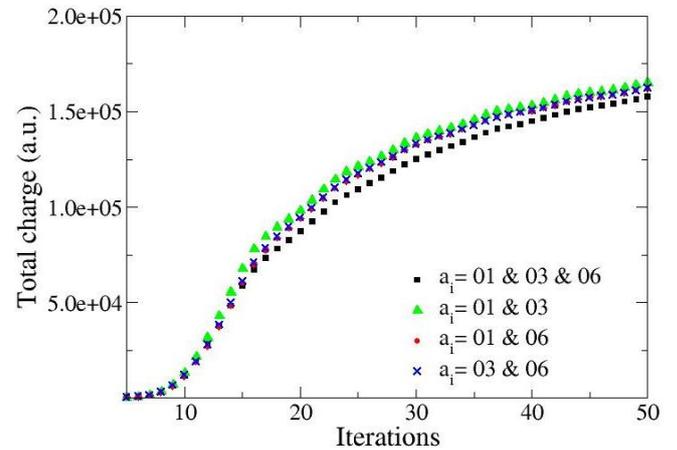

FIG. 2. Running media of the total charge in a 60x60 grid. Each curve represents a different combination of the relative AI significance, $a_i$, and is obtained averaging over 20 realizations. The corresponding mutants are reported in Table II.

In agreement with Eq. (2a), the fastest growth is observed for the mutant-equivalent deprived of HAI-1 and the slowest for the wild type.

This means that the amount of AI expressed by bacteria ($a_i$) determines how fast the growth is: HAI-1 maximally slows it down while CAI-1 maximally enhances it. These results agree, at least qualitatively, with the growth rate data measured for the wild type BB120 and the double-components mutants of *V. harveyi* grown in liquid cultures [6]. As a matter of facts, the detected growth rates are ordered as follows HAI-1⁻ > AI-2⁻> CAI-1⁻ > wild type. Notice the different role of CAI-1 with respect to bioluminescence. Although it has a poor role in bioluminescence emission, it speeds up the growth. This is in line with the source-sink energy dynamics postulated in [6] for the transfer of energy from bioluminescence to cell growth. This suggests that mutants containing CAI-1 exploit energy for colony growth more than for luminescence production. In other words, CAI-1 is cooperative in cell production and behaves as a cheater in bioluminescence. The opposite behavior is observed in AI-1, therefore switching on/off the genes that produce these AIs should help the colony to face different environmental conditions, i.e. the amount of available food. The slowest rate of growth is associated with the highest luminescence, hence the latter should be activated as a protection mechanism against predators, which should be more effective when the colony is struggling to grow. In figure 3(a) a typical colony evolution at two different stages of development (iterations) is reported for the wild type and two-component mutants. Black pixels are used for occupied sites on a 50x50 grid.

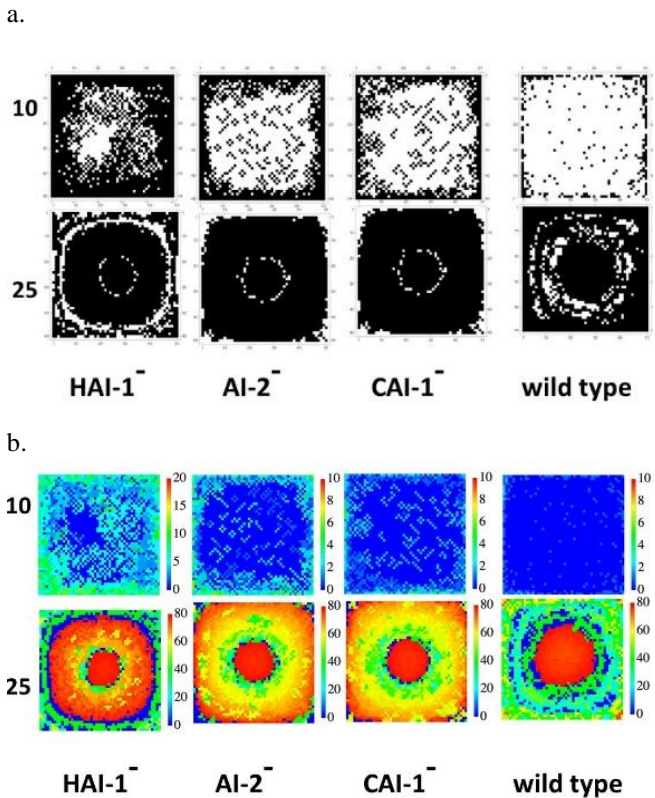

FIG. 3. (a) Screenshots of the different mutant-equivalents (liquid culture) taken at iteration 10 and 25. A. black pixels correspond to occupied sites, white to empty sites. (b) Colors from blue to red refer to increasing values of the charge Q.

The distribution of the charges inside the network for the same 4 mutant-equivalents is represented in figure 3(b). We remark that ring patterns are visible, similar to what happens in colonies of *B.Subtilis* [28] and *E.coli* [29]. In both cases, the origin of this kind of patterns seems to be related to the features of the substrate and the concentration of nutrients [30]. Specifically, these patterns were induced in strains of *E.coli* modified in such a way to mimic a QS system similar to that implemented by *V. harveyi* [Liu-science. Lu-chembio], here used to inhibit cell motility at high density.

The growth of bacteria is affected by several factors like temperature, pH, culture ground, availability of nutrients, and also the type of substrate [28]. Noteworthy, the development of ring-like pattern has been reported to be associated with two characteristic behaviors of a growing colony, namely, consolidation and migration. Under favorable conditions, a concentric ring-like pattern has been associated with the alternating fast expansion and slow growth state [28].

All these factors should be taken in account when fine-tuning the model parameters. In particular, differences appear between the development of the colony in liquid culture [6] compared to what is observed on the substrate of hydroxyapatite.

### C. QS and colony growth on the hydroxyapatite substrate

Concerning the colony growth, growth experiments lasting for 48 h reflect the cumulative effect of different growth rates and regimes because for sufficiently long time the biofilm associated with any mutant is mature and substrate coverage is steady.

As detailed in section III, four different strains of *V. harveyi* were investigated, specifically the wild strain BB120 and its isogenic derivatives that retain double component mutants (HAI-1-, AI-2-, CAI-1-). Regarding this, figures 4a and 4b show stereomicroscopy images of the biomass associated with BB120, JMH603, JAF633 and KM387 grown onto hydroxyapatite substrates when the growth is delayed from 24 to 48 hours, respectively.

In the case of the mature biofilm, OD(595 nm) and CFU/mL data were found to indicate a common trend in the biomass development [25]. As Figure 5 shows, the minimal biofilm development was observed in the case of the wild type strain BB120 and the mutant KM387 (AI-2-). The maximum coverage was associated with JMH603 (CAI-1-) followed by JAF633 (HAI-1-) [25]. Visual inspection of the stereomicroscopy images associated with growth lasting 24 hours (Figure 4a) clearly shows the formation of localized colonies (darker violet spot-like features) with larger density and size in the case of JMH 603. While the spatial development of the biofilm associated with BB120 and KM387 is ring-like with decreasing coverage on going from the periphery to the center of the hydroxyapatite disc, JAF633 and JMH603 exhibit more uniform coverage and the occurrence of localized larger foreground colonies dispersed over a growing background. The evolution of coverage can be clearly observed by comparing the stereomicroscopy images at 24 and 48 hours (Figure 4b).

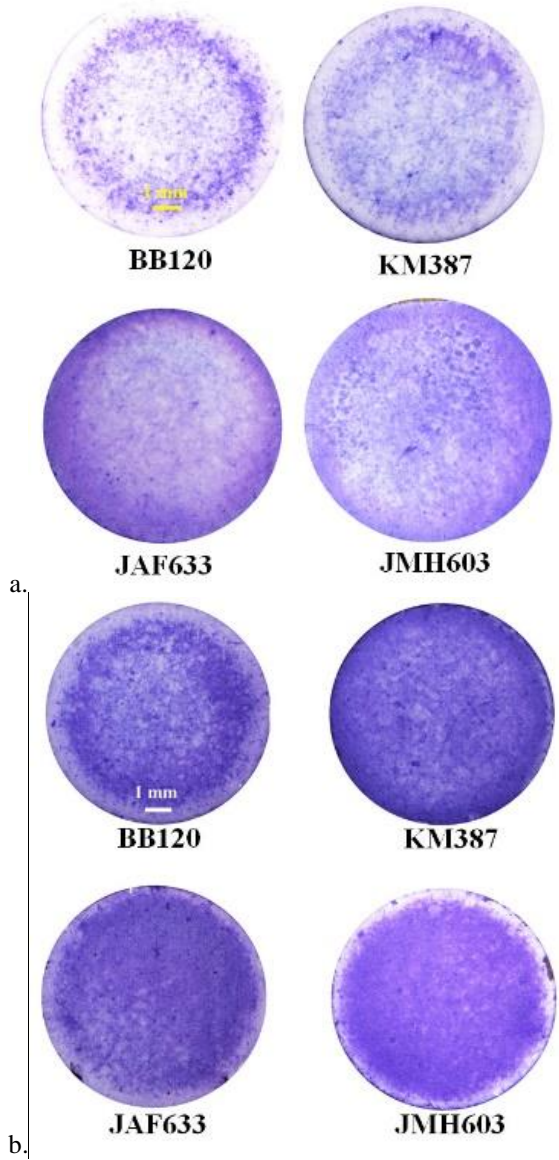

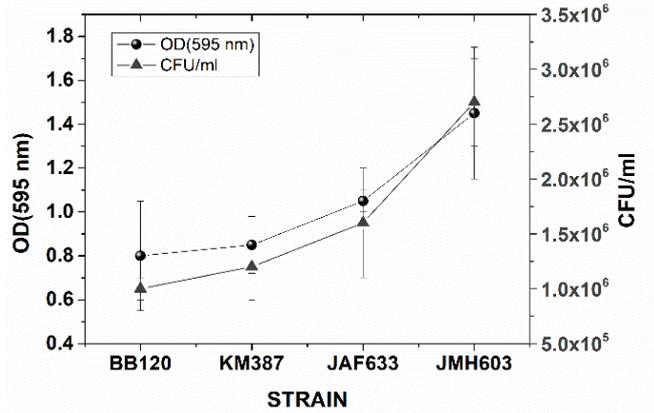

FIG. 5. Comparison between the optical density signal at 595 nm (OD(595 nm)) and CFU/ml counts associated with the four strain BB120, KM387, JAF633 and JMH603 grown onto hydroxyapatite substrates for 48 hours.

TABLE I. Results of the estimation of the relative coverage ($cvg_{rel}$) after 24h and 48h by segmentation of the stereomicroscopy images acquired for the four strains BB120, KM387, JAF 633 and JMH603 grown onto HA discs for 24 and 48 hours.

| Strain | 24h-$cvg_{rel}$ (%) | 48h-$cvg_{rel}$ (%) |
|---|---|---|
| BB120  | 1%   | 1%   |
| KM387  | +6%  | +2%  |
| JAF633 | +15% | +4%  |
| JMH603 | +23% | +16% |

FIG: 4. Images acquired by stereomicroscopy of mature biofilms grown for (a) 24 hours and (b) 48 hours onto hydroxyapatite substrates of the wild type strain BB120 and its isogenic derivatives KM387, JAF633 and JMH603. The violet-color is due to staining of the biomass by the crystal-violet dye.

Based on digital processing (thresholding and segmentation) of the images acquired by stereomicroscopy following growth for 24 and 48 hours, the relative change of coverage of mutants with respect to the wild type strain was calculated, as reported in section III. The obtained estimations, referred to as *24h-cvgrel* and *48h-cvgrel*, are reported in Table I. Relative coverage values systematically increase on going from BB120 to JMH603 through KM387 and JAF633. Such increase is more relevant for growth lasting 24 hours (+6 % for KM387, + 15 % for JAF633 and +23% for JMH603) than for growth lasting 48 hours (+2 % for KM387, + 4 % for JAF633 and +16 % for JMH603) [25].

## IV. DISCUSSION AND CONCLUSIONS

The stochastic model we have proposed describes QS by means a long-range potential produced by charges representing the bacterium strength, i.e., the number of bacteria in an assigned site. The system evolves to maximize its electrostatic energy. This is possible by extracting energy/food from the environment, allowing the system to reach a maximal energy state. In the meantime, the colony develops through several reproductive processes and repeated death-birth events. The role of AIs as mediators of the bacterium-bacterium interaction is expressed by the probability of increasing the value of each charge and is made visible through a percolative mechanism: in the network hosting the colony, channels of electrical current are open between different sites when one of them promotes the charge growth of the other. In this description, QS is monitored by the (fictitious) current flowing in the network.

QS develops inside the colony depending on the amount and distribution of the charges and the complex dynamics of several different processes. A mixed population of cooperative and non-cooperative AIs competing for the public goods seems to underlie the different manifestations of QS [31,32]. As an example, we analyze data coming from different mutants of *V. harveyi* in a liquid culture [6] , which can be described like a well mixed system. In this case, the bioluminescence is mainly sustained by two of the three AIs, i.e., HAI-1 and AI-2, with CAI-1 being a cheater in this respect. On the other hand, the mutants using the CAI-1 channel show a fast colony formation, from which they benefit by increasing their number. These mutants have a low cost in bioluminescence and great reproductive benefit .

This observation could be translated into terms of cooperative coefficients for each of the AIs. Specifically, we assume that the Hill number in Eq.(6) is given by the linear combination:

$$\gamma = \sum_{i=1}^{N} \gamma_i = \sum_{i=1}^{N} c_i a_i \qquad (8)$$

with $N$ being the number of AIs that represent the specific mutant, and $c_i$ the associated cooperative coefficient. Therefore, the values of $\gamma$ used to fit experiments performed on liquid cultures can be associated to the following coefficients: $c_1 = 0$ for CAI-1, $c_2 = \frac{6}{5}$ for AI-2 and $c_3 = \frac{5}{3}$ for HAI-1. In other words, the role of CAI-1 is to activate the partner AI in the mutant, although it does nothing itself, or better, it does not use energy in producing light. In the wild type, all the AIs are present and interact in pairs or all together, for a total of 4 different combinations. As a result, the value of the cooperative coefficient increases and the choice: $c_1 = 0$ for CAI-1, $c_2 = 4$ for AI-2 and $c_3 = 4$ for HAI-1 , i.e. γ =3.6 reproduces experimental data.

As we remarked, several investigations report that environmental conditions, food availability or a specific substrate [28,32] affect the colony growth as well as bioluminescence [1]. Therefore, we also analyzed the colony development on a solid substrate, specifically, hydroxyapatite. In this case, mutants using the CAI-1 channel have the slowest growth.

We suggest that these findings can be accommodated in our model assuming that the 3 AIs are produced in different amount (AI significance, $a_i$) on HA substrate with respect the liquid culture [1]. In particular, the measured rate of growth suggests that, on HA, CAI-1 is produced in larger amount with respect the other two AIs, so that it is the production itself which drains energy from growth. As a matter of facts, using $a_1 = 0.6$ for CAI-1, $a_2 = 0.1$ for AI-2 and $a_3 = 0.3$ for HAI-1, the results shown in figure 2 agree qualitatively with those of the colony grown on HA ( see also Table II).

Furthermore, by using the cooperative coefficients previously introduced, and we can calculate the bioluminescence for mutants grown on this kind of support. The results are reported in Figure 1 (inset), using a 50x50 grid. Finally, to compare these results with those referring to the liquid culture, data were normalized to the wild type value of the liquid culture. These results suggest that except for the wild type, the other mutants are generally dim (the most luminescent gives a response of about 1% of the wild type). As a final remark, we note that it has been observed [33] that *V. campbellii* is less bright than *V. harveyi*, even in liquid cultures, therefore, we cannot say whether our data regarding bioluminescence and colony growth are strictly due to the specific substrate or are a peculiar behavior of this strain. Further investigations are underway on this point.

In conclusion, our model predicts that the QS is at the origin of the bioluminescence and of the development of the colony because for both phenomena it carries the quantity of charge that is distributed in the colony. Furthermore, bioluminescence depends on the cooperativity between different AIs and the amount of AIs, and both *V. harveyi* and *V. campbelli* could use it as a defense mechanism to protect the colony when growth is not very rapid.

TABLE II. Correspondence between the mutants and the different combinations of AI significance ($a_i$).

| $a_i$ combinations | *Liquid culture* | *Hydroxyapatite* |
|---|---|---|
| 010306 | Wild type | Wild type |
| 0103 | HAI-1[-] | CAI-1[-] |
| 0106 | AI-2[-] | HAI-1[-] |
| 0306 | CAI-1[-] | AI-2[-] |

### ACKNOWLEDGMENTS

M. C. gratefully acknowledges Prof. Pietro Alifano and Dr. Matteo Calcagnile (Di.S.Te.B.A.- University of Salento- Lecce, Italy) for providing bacteria strains and hydroxyapatite substrates, preparing bacteria coltures and crystal violet staining.

Funding: M.C. was supported by the BANDO POR PUGLIA FESR-FSE 2014 / 2020 - Research for Innovation (REFIN)-Regione Puglia- Proposal Number: 012C1187. M.B. was supported by INFN, Iniziativa Specifica GSS.

1. J. M. Henke and B. L. Bassler, J. Bacteriol. **186**(20), 6902-6914 (2004).
2. M. B. Miller and B. L. Bassler, Quorum sensing in bacteria. Annu. Rev. Microbiol., **55**(1), 165-199 (2001).
3. B. Striednig, and H. Hilbi, H. Trends Microbiol., **30** (4), 379-389 (2022).
4. Y.V. Zaitseva, A. A. Popova and I. A. Khmel, Russ. J. Genet. **50**, 323-340 (2014).

5. L. Del Gigia-Aguirre, W. Sánchez-Yebra-Romera, S. García-Muñoz, and M. Rodríguez-Maresca, New microbes and new infections, **19,** 15 (2017).
6. Z. E. Nackerdien, A. Keynan, B. L. Bassler, J. Lederberg, and D. S. Thaler, Plos one, **3**(2), e1671 (2008).
7. T. Varadavenkatesan, R. Vinayagam, S. Pai, B. Kathirvel, A. Pugazhendhi, and R. Selvaraj, Prog. Org. Coat., **151**, 106056 (2021).
8. J. D. Dockery, and J. P. Keener, Bull. Math. Biol., **63**(1), 95-116 (2001).
9. J. P. Ward, J. R. King, A. J. Koerber, P. Williams, J. M. Croft and R. E. Sockett, Math. Med. Biol., **18**(3), 263-292 (2001).
10. J. Pérez-Velázquez, M. Gölgeli, and R. García-Contreras, R. Mathematical modelling of bacterial quorum sensing: a review. Bull. Math. Biol., **78**, 1585-1639 (2016).
11. A. B. Goryachev, Chemical reviews, **111**(1), 238-250 (2011).
12. M. Weber, and J. Buceta, BMC Syst. Biol., **7**(1), 1-15 (2013).
13. P. Sinclair, C. A. Brackley, M. Carballo-Pacheco, and R. J. Allen, PRL, **129**(19), 198102 (2022).
14. M. R. Frederick, C. Kuttler, B. A. Hense, and H. J. Eberl, Theor. Biol. Medical Model. **8**, 1-29 (2011).
15. R. D. Acemel, F. Govantes, and A. Cuetos, Computer simulation study of early bacterial biofilm development. Sci. Rep. **8**(1), 5340 (2018).
16. E. Ben-Jacob, O. Schochet, A. Tenenbaum, I. Cohen, A. Czirok, A. and T. Vicsek, Nature, **368**(6466), 46-49 (1994).
17. J. Pérez-Velázquez, B. Quiñones, B. A. Hense and C. Kuttler, Ecol. Complex, **21**, 128-141 (2015).
18. B. A. Hense, C. Kuttler, J. Müller, M. Rothballer, A. Hartmann, and J. U. Kreft, Nat. Rev. Microbiol., **5**(3), 230-239 (2007).
19. E. Alfinito, M. Cesaria and M. Beccaria, Biophysica, **2**(3), 281-291 (2022).
20. E. Alfinito, J. Pousset, and L. Reggiani, *Proteotronics: development of protein-based electronics* (CRC Press, Boca-Raton, 2015).
21. G. Craciun, M. D. Johnston, G. Szederkényi, E. Tonello, J. Tóth, and P. Y. Yu, MBE , 17(1), 862-892 (2020).
22. A. V. Hill, J. Physiol. **40**: iv–vii 1910.
23. S. Goutelle, M. Maurin, F. Rougier, X. Barbaut, L. Bourguignon, M. Ducher and P. Maire, Fundam. Clin. Pharmacol., **22**(6), 633-648. (2008).
24. B. Lin, Z. Wang, A. P. Malanoski, E. A. O'Grady, C. F. Wimpee, V. Vuddhakul, A. Jr Nelson, F. L. Thompson, B. Gomez-Gil, and G. J. Vora, Environ. Microbiol. Rep., **2**(1), 81-89 (2010).
25. M. Cesaria, M. Calcagnile, P. Alifano, and R. Cataldo, Int. J. Mol. Sci., **24**(6), 5423 (2023).
26. E. Peeters, J. Microbiol. Methods, 72 (2), 157-165 (2008).
27. C. A. Schneider, W. S. Rasband, and K. W. Eliceiri, Nat. Methods, **9** (7), 671-675 (2012).
28. H. Shimada, T. Ikeda, J. I. Wakita, H. Itoh, S. Kurosu, F. Hiramatsu, M. Nakatsuchi, Y. Yamazaki, T. Matsuyama and M. Matsushita , J. Phys. Soc. Japan, **73**(4), 1082-1089 (2004).
29. C. Liu, X. Fu, L. Liu, X. Ren, C. K. Chau, S. Li, ... and J. D. Huang, Science, **334**(6053), 238-241 (2011).
30. J. Lu, E. Şimşek, A. Silver, and L. You, Curr. . Opin. Chem. Biol., **68**, 102147 (2022).
31. E. L. Bruger, D. J. Snyder, V. S. Cooper, V. S. and C.M. Waters, C. M. ISME J., **15**(4), 1236-1247 (2021).
32. H. Monaco, K. S. Liu, T. Sereno, M. Deforet, B. P. Taylor, Y. Chen, C. C. Reagor, and J. B. Xavier, Nat. Commun., **13**(1), 721 (2022).
33. T. Defoirdt, W. Verstraete, and P. Bossier, J. Appl. Microbiol. **104**(5), 1480-1487 (2008).